\documentclass[final,1p,times]{elsarticle} 

\usepackage{graphicx}
\usepackage{amssymb} 
\usepackage{amsthm} 
\usepackage{lineno}

\journal{Nuclear Physics A} 

\begin{document}

\begin{frontmatter} 

\title{Thermodynamics of the QCD transition from lattice}

\author{Szabolcs Bors\'anyi (for the Wuppertal-Budapest\fnref{WBC} Collaboration)}
\fntext[WBC] {borsanyi@uni-wuppertal.de}
\address{Theoretische Physik, Bergische Universit\"at Wupperal, D-42119, Germany}


\begin{abstract} 
    We discuss recent developments in lattice QCD for the
    bulk thermodynamics of the transition.
    We review the current status of the equation of state, 
    the possible implications of a magnetic field and the fluctuations
    of conserved charges, like the net electric charge or baryon number.
    We also show predictions for higher cumulants, which will be
    experimentally available in near future.
\end{abstract} 

\end{frontmatter} 


\section{Introduction}

Lattice simulations aiming to describe the features of quark gluon plasma and
its transition to confined matter have become a field of high precision
in the past years. An important goal of the efforts
on the lattice is to provide theoretical background for collision
experiments. The relevance of lattice results are most prominent
for the heavy ion physics pursued at RHIC and LHC, where matter with
high energy density, but moderate or small chemical potentials are produced.

Lattice simulations solve Quantum Chromodynamics (QCD) in equilibrium. The
solution emerges from a continuum limit, which is an extrapolation from a set
of simulations on lattices with smaller and smaller lattice spacings ($a$).
Finite temperature ($T$) is naturally implemented by a torus in Euclidean time.
In most thermodynamics studies the lattice spacing is expressed through the
number of lattice points in the Euclidean time direction: $N_t$, which
translates to $a=1/(N_tT)$. In most early papers $N_t$ was set to $4$ or $6$.
By today $N_t=16$ data sets are not uncommon, and are believed to be necessary
for a controlled continuum extrapolation even with improved actions. 

Longer correlations (inverse pion mass) require larger computer resources to
simulate the theory. In the past decade lattice field theory has seen a great
development in gradually reducing the quark masses down to their physical
values \cite{Hoelbling:2011kk}. Simultaneously to this trend our increasing
understanding of various discretization effects has lead to more efficient
discretization schemes (actions) that allow a continuum extrapolation at an
affordable cost \cite{Morningstar:2003gk,Follana:2006rc}.
In view of these developments here we only discuss results
with  physical  (or nearly physical) quark masses close to or in the continuum limit.

\section{Transition temperature}

Evidence form lattice QCD shows that at zero chemical potential
the transition from the chirally broken, confining phase of hadrons
to the quark gluon plasma phase is a crossover
\cite{Aoki:2006we}.

The transition temperature of a crossover transition depends on the observable
one chooses. At
vanishing light quark masses the chiral condensate is a genuine order
parameter, which can be used to extract a chiral transition temperature. At
sufficiently high quark masses the deconfinement transition is identified
by a discontinuity in the Polyakov loop as a function of temperature.
Varying the quark mass in the space of theories the relevance of chiral and
deconfinement aspects can be tuned. 
With sufficiently light $u$ and $d$ quarks an O(4) scaling of the chiral
observables is conjectured \cite{Ejiri:2009ac,Kaczmarek:2011zz}.  This universal
mass dependence has been used to relate the transition temperature ($T_c$) in
QCD to $T_c$ for the actual simulation parameters \cite{Bazavov:2011nk}. 

The Wuppertal-Budapest collaboration has simulated the QCD transition with
physical quark masses in the stout staggered formulation
\cite{Aoki:2005vt}. The lattice resolutions range from $N_t=6$ to $N_t=16$.
From the inflection point of the chiral condensate we find $T_c=155(3)(3)$~MeV,
where the first of the errors is statistical, the second is systematic,
including the uncertainties of the scale setting \cite{tc-onetwo,tc-three}.
Other $T_c$ definitions from chiral observables span a range of $147-157$ MeV.
Recently the HotQCD collaboration has published a transition
temperature, $154(8)(1)$~MeV, based on the O(4) scaling \cite{Bazavov:2011nk},
which is very well compatible with the range given by the Wuppertal-Budapest
collaboration in 2006. This is in contrast to the earlier continuum result 
(192(4)(7) MeV) by the RBC-Bielefeld group. A key improvement since then
was the adoption of the \texttt{HISQ} action \cite{Follana:2006rc}.

The quoted values of $T_c$ are all based on lattice observables that are
remnants of the chiral order parameter. One can also study other observables,
such as the Polyakov loop,
which is the remnant order parameter of the deconfinement transition.
The Polyakov loop as well as the inflection point is renormalization scheme
dependent.  E.g. the scheme used in Ref.~\cite{tc-onetwo} was different
than in our latest work \cite{tc-three}. We made this change in our
convention to be compatible with Ref.~\cite{Bazavov:2009zn} in every
detail. While the comparison of our result to the \texttt{asqtad}
data has showed a disagreement, the new \texttt{HISQ} data of the HotQCD
collaboration was in agreement with ours \cite{Bazavov:2011nk}.

\begin{figure}[htbp]
\begin{center}
\includegraphics[height=0.2\textheight]{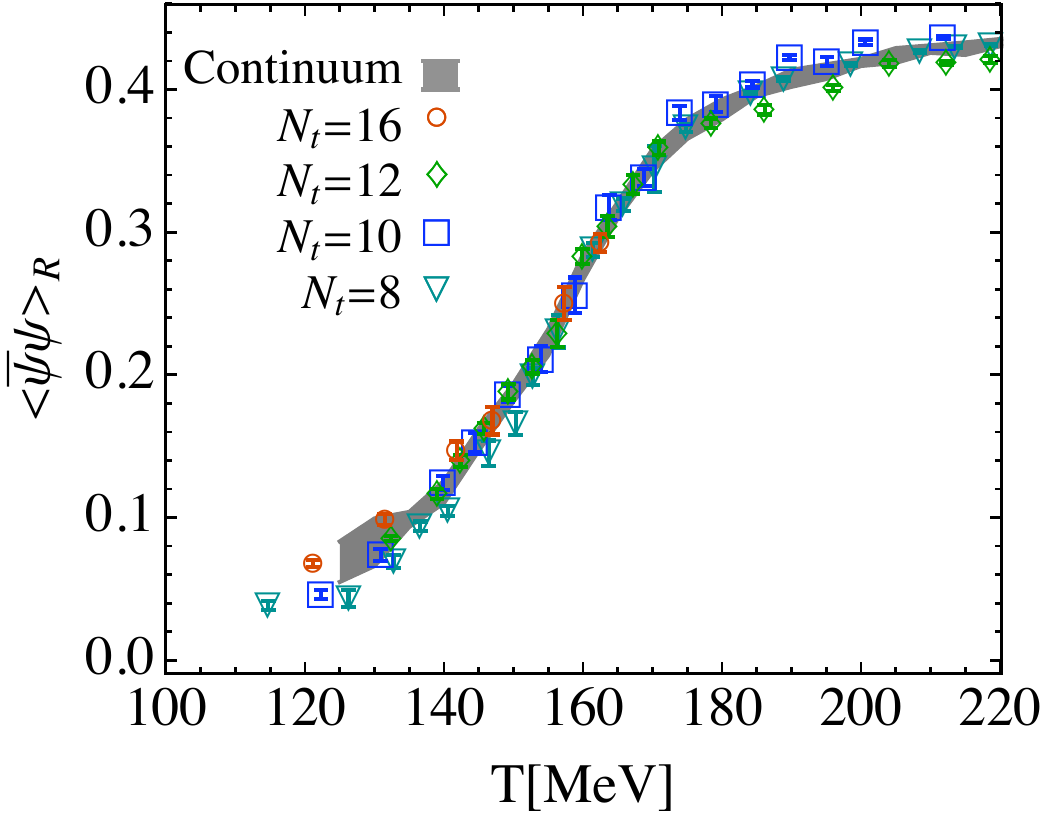}
\hspace{0.1\textwidth}
\includegraphics[height=0.2\textheight]{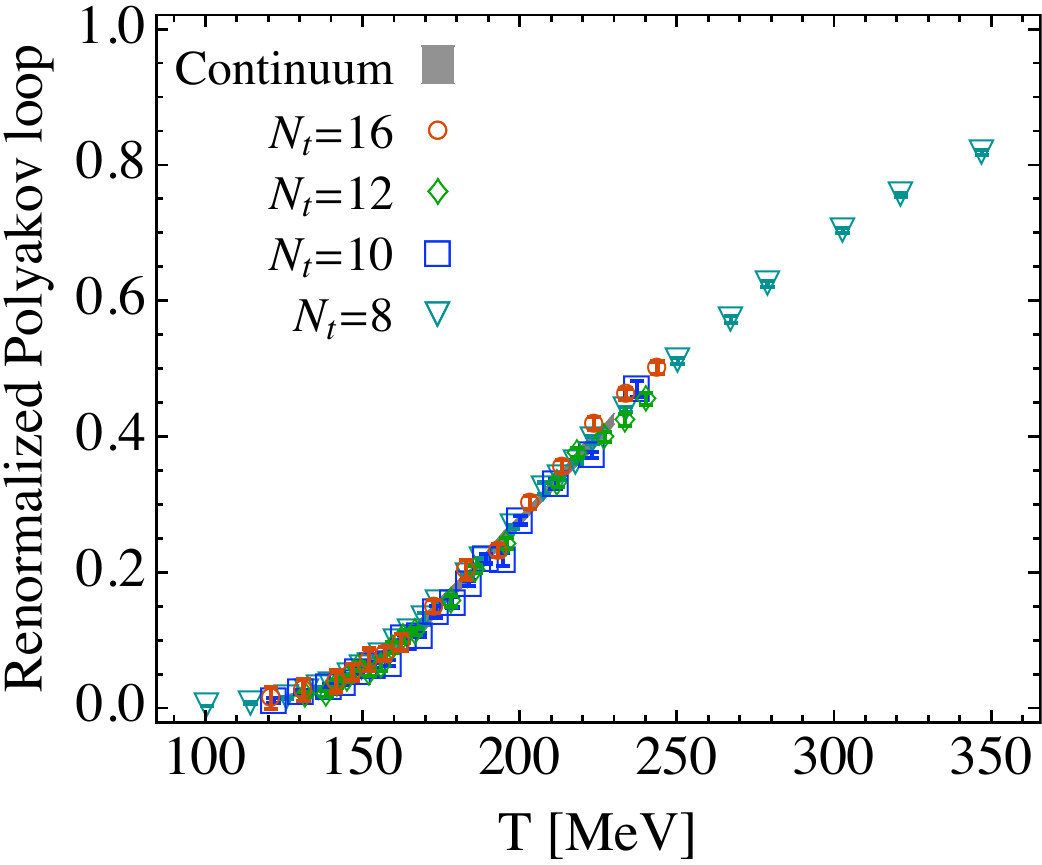}
\end{center}
\vspace*{-3mm}
\caption{
\textit{Left:} The renormalized chiral condensate. We used the inflection point
to define a transition temperature: 155(3)(3) MeV. \textit{Right:} the renormalized
Polyakov loop. Although one can precisely measure its temperature dependence,
one cannot easily pinpoint an inflection point. The shape of the Polyakov
loop curve, as well as any derived transition temperature depends on the
details of the renormalization scheme \cite{tc-onetwo,tc-three}. 
}
\label{fig:tc}
\end{figure}

There are other observables that indicate deconfinement. The normalized
energy density $\epsilon/T^4$ has an inflection point at $157(4)(3)$ MeV,
for the trace anomaly we have $154(4)(3)$ MeV. The speed of sound has
a minimum at around $\sim 145(5)$ MeV. These temperatures cover roughly
the same range as we find from chiral observables. From the strange
susceptibility, which characterizes the deconfinement of the strange
degree of freedom we have a somewhat higher inflection point. Of course,
in a broad crossover the definition of the most singular point is also
ambiguous.

The concept of an identical chiral and deconfinement transition temperature
was promoted by early lattice data \cite{Karsch:2001cy}, that positioned
the peak of the chiral and the Polyakov loop susceptibilities at the same
temperature: $T_c$.  Lacking resources, only theories with heavier quarks 
were simulated at the time. Results with somewhat smaller quark
masses on coarse lattices confirmed this picture \cite{Cheng:2007jq}.
Solutions to two-flavor QCD based on the functional renormalization group have
shown the same conclusion \cite{Braun:2009gm}.  In full (2+1 flavor) QCD,
however, we see a very broad transition. Thus the transition temperatures based
on various observables generally do not coincide, they have their values in
this broad transition region.

\begin{figure}[t]
\begin{center}
\includegraphics[height=0.2\textheight]{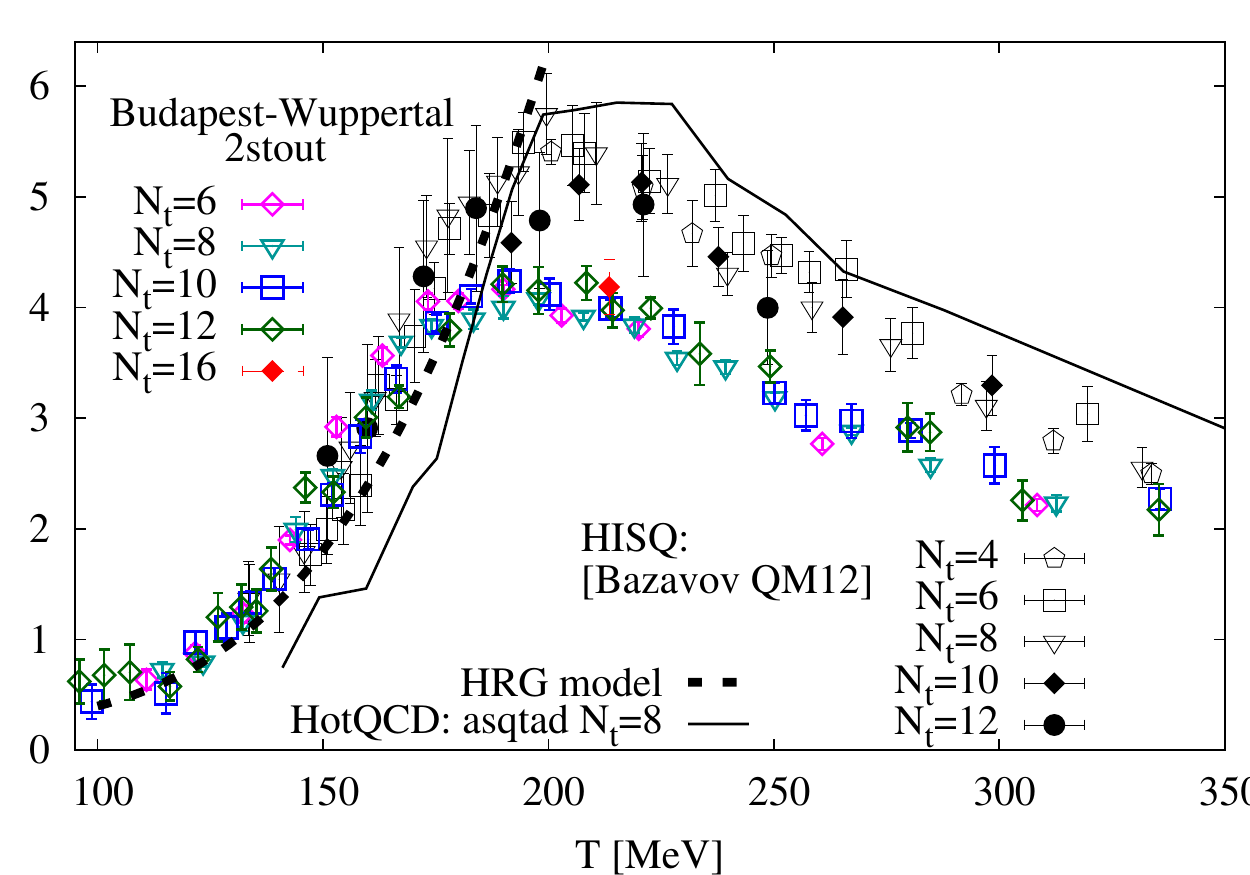}
\hspace{0.05\textwidth}
\includegraphics[height=0.2\textheight]{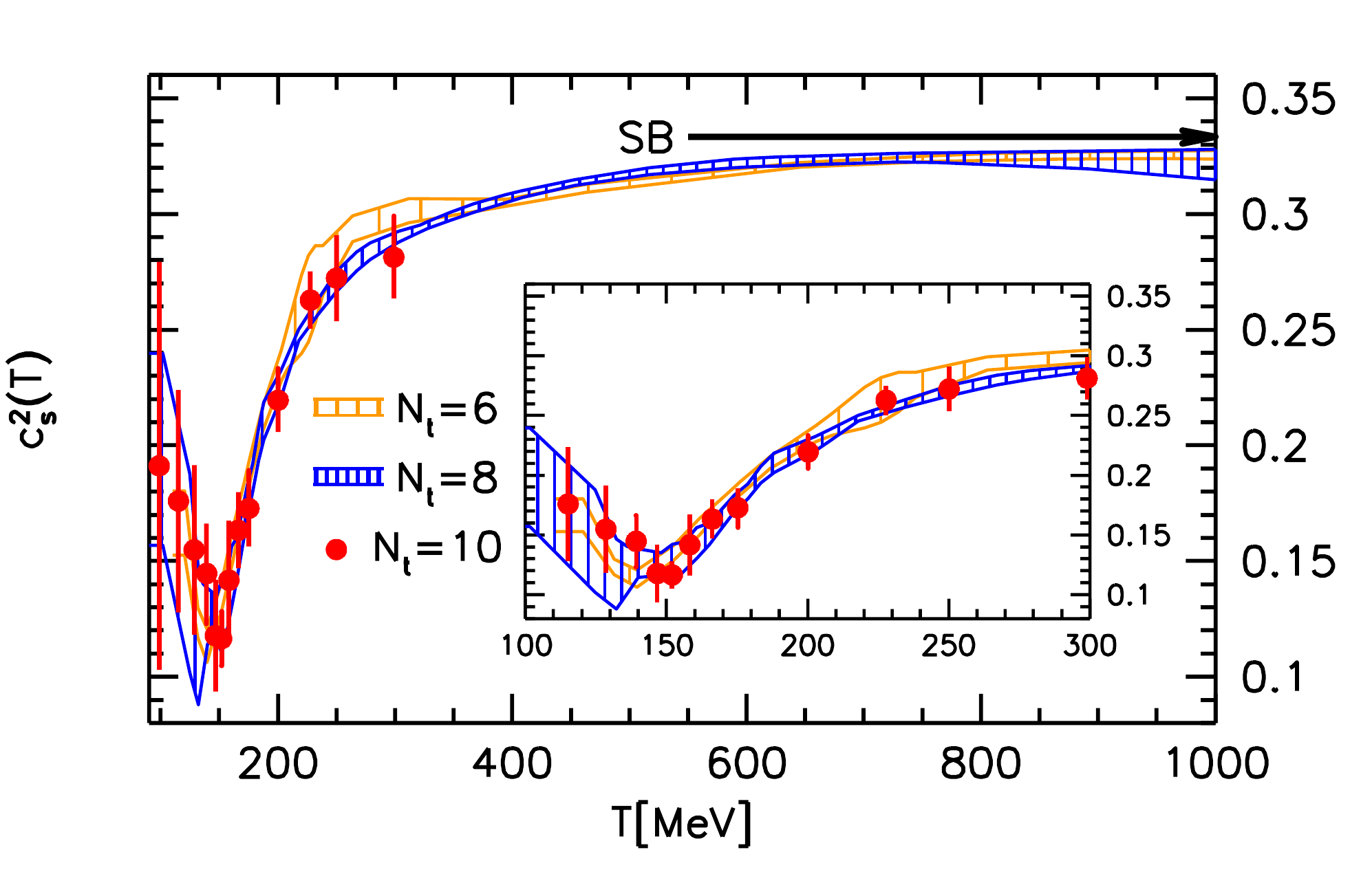}
\end{center}
\vspace*{-3mm}
\caption{
\textit{Left:} 
The trace anomaly at several lattice spacings, either with
\texttt{2stout} action of the Wuppertal-Budapest collaboration
\cite{Borsanyi:2010cj,ClaudiaQM12}, or with the \texttt{HISQ} action of the HotQCD
collaboration \cite{AlexeiQM12}. For comparison we show the Hadron Gas
Resonance model's prediction as well as the historic \texttt{asqtad} data of
Ref.~\cite{Bazavov:2009zn}. The use of the \texttt{HISQ} action has significantly reduced the
discrepancy between the two collaborations.  \textit{Right:}
The speed of sound by the Wuppertal-Budapest collaboration at three
lattice spacings \cite{Borsanyi:2010cj} (these data are not yet available from \texttt{HISQ} calculations).
}
\label{fig:eos}
\end{figure}

\section{Equation of state}

The pressure and energy density of the quark gluon plasma is one of the key
targets of lattice QCD simulations. These form the equation of state, which is
required to close the equations of relativistic hydrodynamics
\cite{Kolb:2003dz}.

Early works on the QCD equation of state were dominated by quenched results
\cite{Boyd:1996bx,Okamoto:1999hi}. These were recently extended in range and
precision, so that a connection to perturbation theory could be made
\cite{Borsanyi:2012ve}. The calculations were also extended towards larger gauge groups
\cite{Lucini:2012gg}.  For several years we have seen various exploratory
studies with non-physical or not renormalized quark masses. Since
Ref.~\cite{Aoki:2005vt} physical simulations have been feasible. In staggered
simulations discretization effects raise the mass of the hadrons, and through
this $T_c$ appears to be higher, and the energy and entropy density function is
found steeper than physical, and the trace anomaly shows a higher peak. This is
an artefact (``taste-breaking'') which is absent after continuum extrapolation,
its impact on coarse lattices and older actions (like \texttt{asqtad})
was emphasized in Refs.~\cite{tc-onetwo,tc-three,Huovinen:2009yb}.  The
Wuppertal-Budapest collaboration exploited the favourable taste-breaking
features of the \texttt{stout} action and calculated a continuum estimate of
the equation of state \cite{Borsanyi:2010cj} based on $N_t=6,8$ and $10$ runs.
HotQCD used the \texttt{asqtad} and \texttt{p4} actions at $N_t=6$ and $8$
\cite{Bazavov:2009zn,Cheng:2009zi}, though these were later found to be
incompatible with results with the better \texttt{HISQ} action \cite{AlexeiQM12}.  For a
comprehensive review see Ref.~\cite{Philipsen:2012nu}.

\begin{figure}[htbp]
\begin{center}
\includegraphics[width=0.45\textwidth]{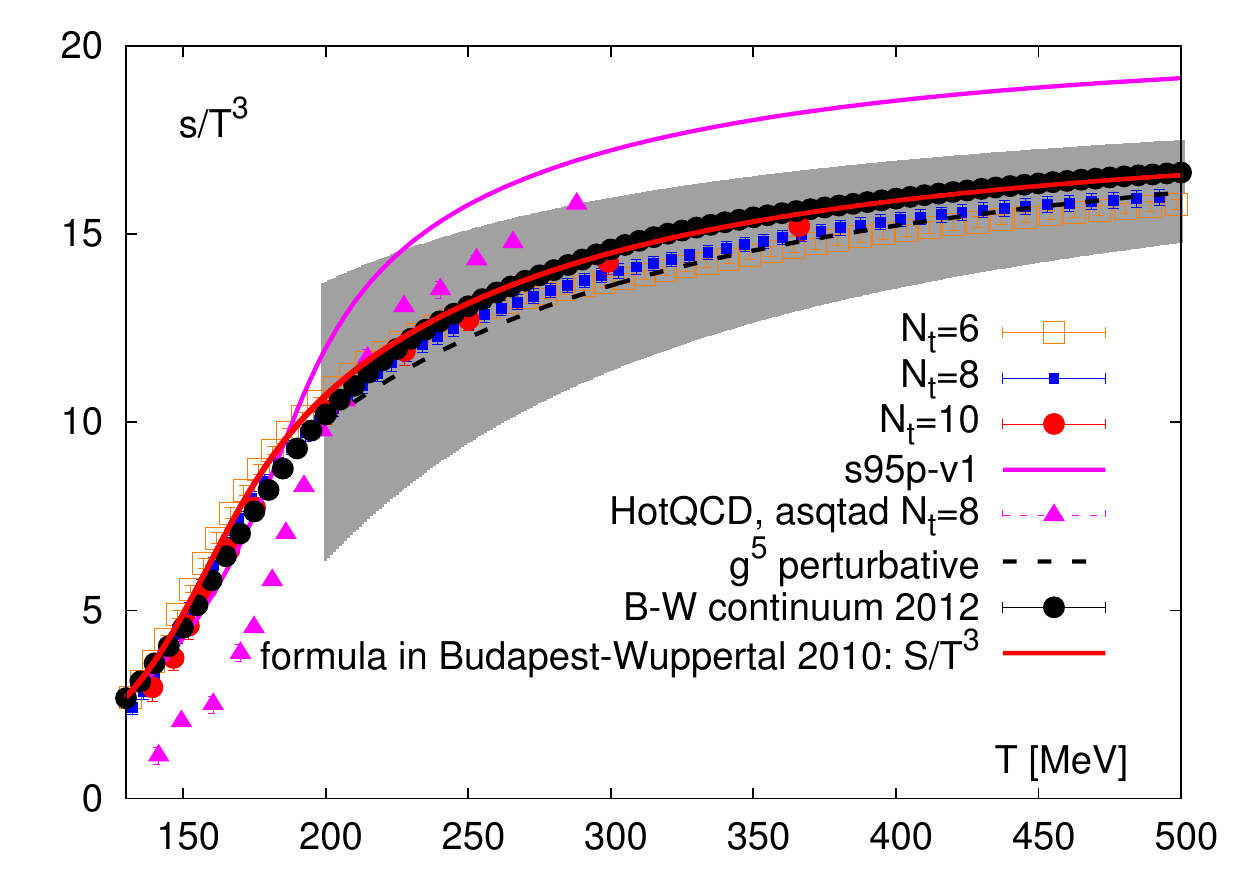}
\hspace{0.05\textwidth}
\includegraphics[width=0.45\textwidth]{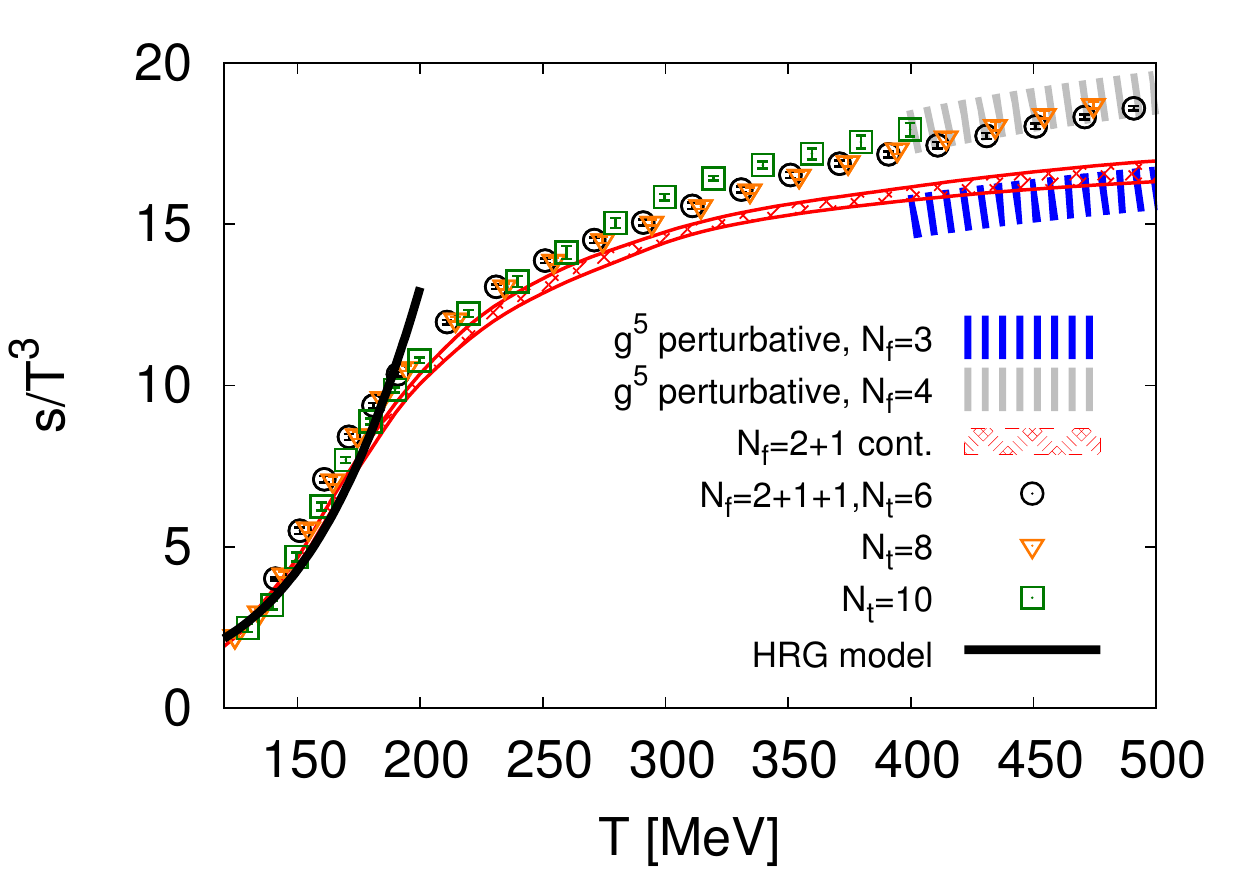}
\end{center}
\vspace*{-2mm}
\caption{
The entropy density as a function of temperature. 
\textit{Left:} We show the lattice data by the Wuppertal-Budapest collaboration
($N_t=6,8$ and $10$) together with the parametrization (solid red line) of
these data sets published in Ref.~\cite{Borsanyi:2010cj}. The black dots
show the result of a recent continuum extrapolation based on increased
statistics and finer lattices ($N_t=12$). We see almost no deviation from
the earlier parametrization. Since there is no \texttt{HISQ} data for the entropy density yet,
we plotted HotQCD's \texttt{asqtad} data set at $N_t=8$ for comparison (magenta
triangles) \cite{Bazavov:2009zn}. As discussed already (Fig.~\ref{fig:eos}),
this set features a higher-than-physical transition temperature and thus it
shows steeper equation of state and higher peak in the trace anomaly.  As
remedy, the \texttt{s95p} parametrization was introduced \cite{Huovinen:2009yb}
by replacing lattice data in and below the transition range by the Hadron
Resonance Gas prediction. By this, the slope of the entropy curve was left
unchanged, but the whole curve was shifted up towards the Stefan-Boltzmann
limit.  \textit{Right:} The contribution of the charm sea quark to the entropy density.
The Wuppertal-Budapest collaboration's data with $N_t=6,8$ and $10$ is shown
for dynamical $2+1+1$ flavor simulations together with its charmless continuum
data and the respective perturbative estimates \cite{ClaudiaQM12}.
}
\label{fig:entropy}
\end{figure}

At this conference, we heard two contributions discussing the equation
of state. The Wuppertal-Budapest collaboration has updated its
continuum estimate of 2010 \cite{Borsanyi:2010cj} by adding
a finer dataset ($N_t=12$) and an even finer check point ($N_t=16$)
and using these in a continuum extrapolation \cite{ClaudiaQM12}.
A. Bazavov (HotQCD collaboration) has presented data at $N_t=4,6,8,10$ and
$12$ \cite{AlexeiQM12}. First, the trace anomaly was obtained form the lattice
simulations. Then, the other quantities, like pressure, entropy density or
speed of sound can be derived using thermodynamic relations. The results are
summarized in Fig.~\ref{fig:eos} and \ref{fig:entropy}.  The missing
contribution of the charm quark has been estimated in C.~Ratti's talk, by
actually performing new 2+1+1 flavor simulations (see
Fig.~\ref{fig:entropy}/right). The charm's contribution to the pressure or
energy density was found to be relevant from about 250-300 MeV temperature
\cite{ClaudiaQM12}.

\begin{figure}[htbp]
\begin{center}
\includegraphics[width=0.41\textwidth]{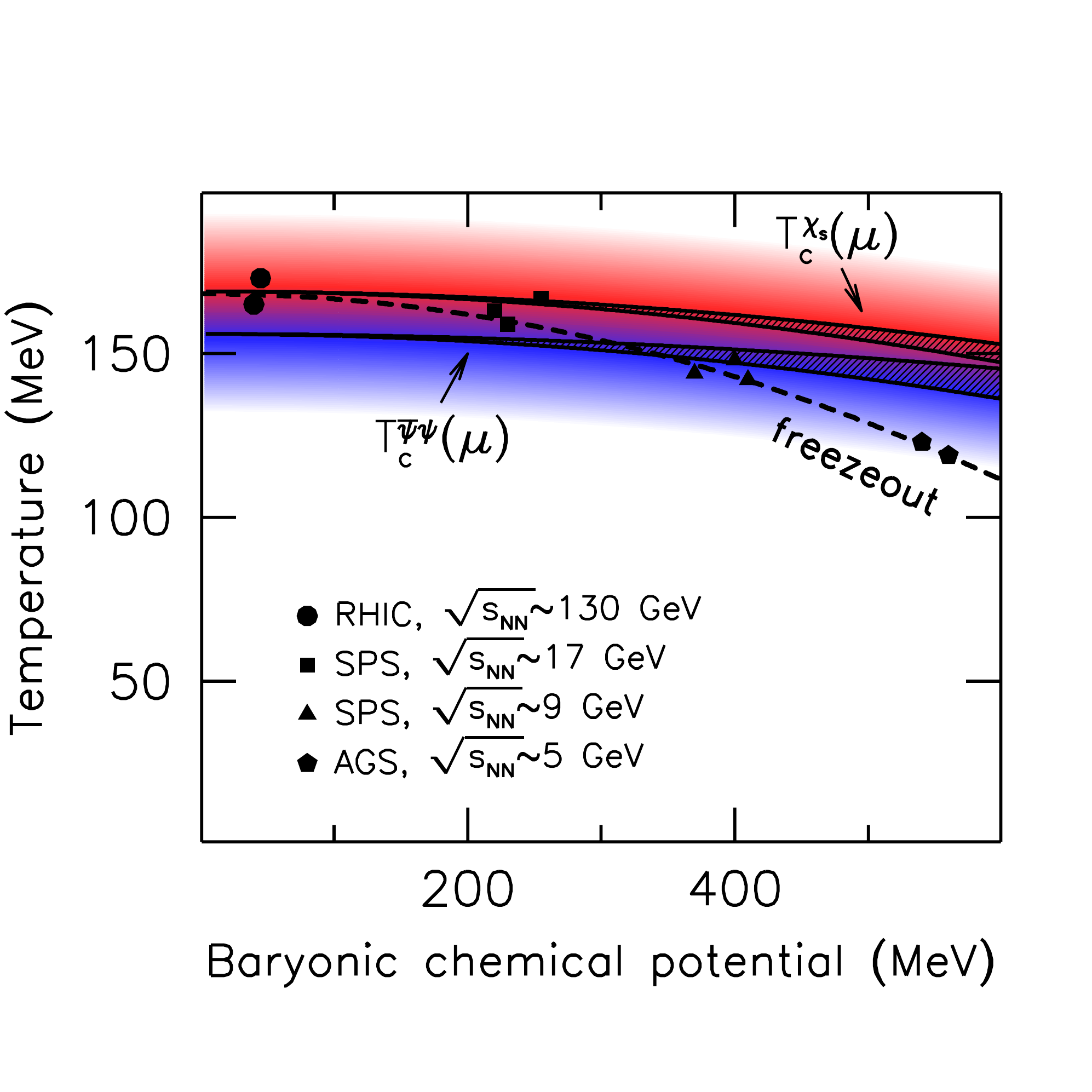}
\hspace{5mm}
\raisebox{4mm}{\includegraphics[width=0.45\textwidth]{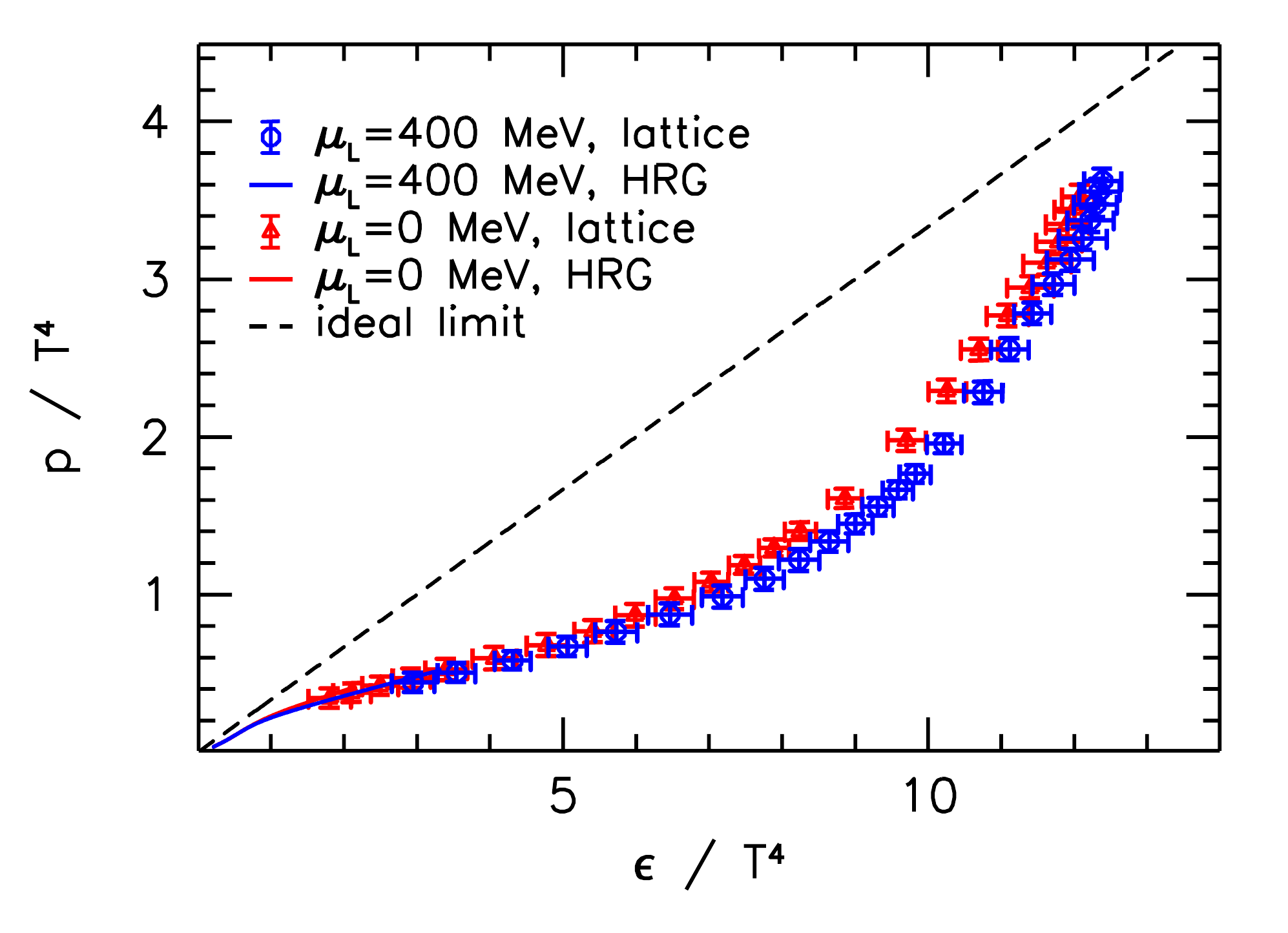}}
\vspace*{-9mm}
\end{center}
\caption{
\textit{Left:}
The QCD phase diagram at small chemical potentials \cite{Endrodi:2011gv}. The
transition temperature is plotted to leading order in $\mu_B^2$.
$T_c^{\bar\psi\psi}$ was extracted from the chiral condensate, and
$T_c^{\chi_s}$ is derived from the strange susceptibility. The solid bands
represent the statistical uncertainty, while the color bands stand for the
measured width of the transition. At this level of the expansion, there is no
indication for the transition getting stronger with increasing $\mu_B$. The
dashed line show the chemical freeze-out curve of Ref.~\cite{Cleymans:1998fq}.
\textit{Right:}
Like $T_c$, the equation of state can also be extrapolated to finite $\mu_B$.
Here $\mu_L$ refers to the baryo-chemical potential while the mean strageness
was kept zero to linear order. For a detailed discussion of these
continuum results and a simple parametrization, see Ref.~\cite{Borsanyi:2012cr}.
}
\label{fig:phasediagram}
\end{figure}

\section{Phase diagram}

The transition line in the $T-\mu_B$ phase diagram at small chemical
potentials is necessarily a line of crossover, with
different width and curvature for various observables.
Direct Monte-Carlo simulations of the four-dimensional quantum field
theory at finite baryo-chemical potential are at present not possible.
Nevertheless, starting with Ref.~\cite{Fodor:2001au} a renewed interest has
been seen for $\mu_B>0$ questions in lattice QCD, and
several indirect strategies exist \cite{Fodor:2009ax,Philipsen:2010gj}.

The curvature of the transition line in full QCD has been found very
small by recent lattice simulations. In Ref.~\cite{Kaczmarek:2011zz}
simulations on $N_t=8$ lattices were used to map the chiral condensate
and its $\mu_B^2$ dependence onto a universal O(4) scaling behaviour as a
function of the quark mass. Evaluating the fits in the physical point
gave $\kappa=0.0066(2)(4)$, with
$\kappa=-T_c\left. d{T_c(\mu_B^2)}/d{\mu_B^2}\right|_{\mu_B=0}$.
The continuum result has been calculated by the Wuppertal-Budapest
collaboration \cite{Endrodi:2011gv}, with simulations explicitly in the
physical point. Their result was $\kappa=0.0066(20)$. In this work,
a similar analysis have been made for the strange susceptibility,
there the curvature was found to be $0.0089(14)$. In addition, the
width of the transition has also been extrapolated to $\mu_B>0$.
This showed that the transition is not getting stronger with increasing
chemical potential, but rather it stays a crossover
(see Fig.~\ref{fig:phasediagram}/left).

This latter conclusion is in line with the work by Philipsen and de Forcrand
claiming the absence of a critical endpoint, at least in 
the range where analytical continuation or Taylor expansion can work
\cite{deForcrand:2006pv}. This result was often confronted
with the critical point found by Fodor and Katz \cite{Fodor:2004nz}. Note,
however, that at small chemical potentials even Ref.~\cite{Fodor:2004nz}
finds a slightly weakening transition, this trend then reverses at higher
$\mu_B$, where the available Taylor expansion from imaginary $\mu_B$ no longer
works. As the authors also emphasize, these results
\cite{deForcrand:2006pv,Fodor:2004nz} came from coarse lattices. A continuum
limit has not yet been feasible (because of the high computational costs), which leaves us without any conclusive lattice
evidence for a critical endpoint or a first order line at high $\mu_B$.

Having no singularity in a large range for chemical potentials, the equation
of state, too, can be extrapolated to $\mu_B>0$.
By integrating the trace anomaly we have $\log Z$ at the $\mu_B=0$.
The baryon number susceptibility $\sim \partial^2\log
Z/(\partial\mu_B)^2|_{\mu_B=0}$ is also accessible from lattice, thus, a linear
expansion is possible. The continuum result was given in Ref.~\cite{Borsanyi:2012cr} (see Fig.~\ref{fig:phasediagram}).

\begin{figure}[htbp]
\begin{center}
\includegraphics[height=0.2\textheight]{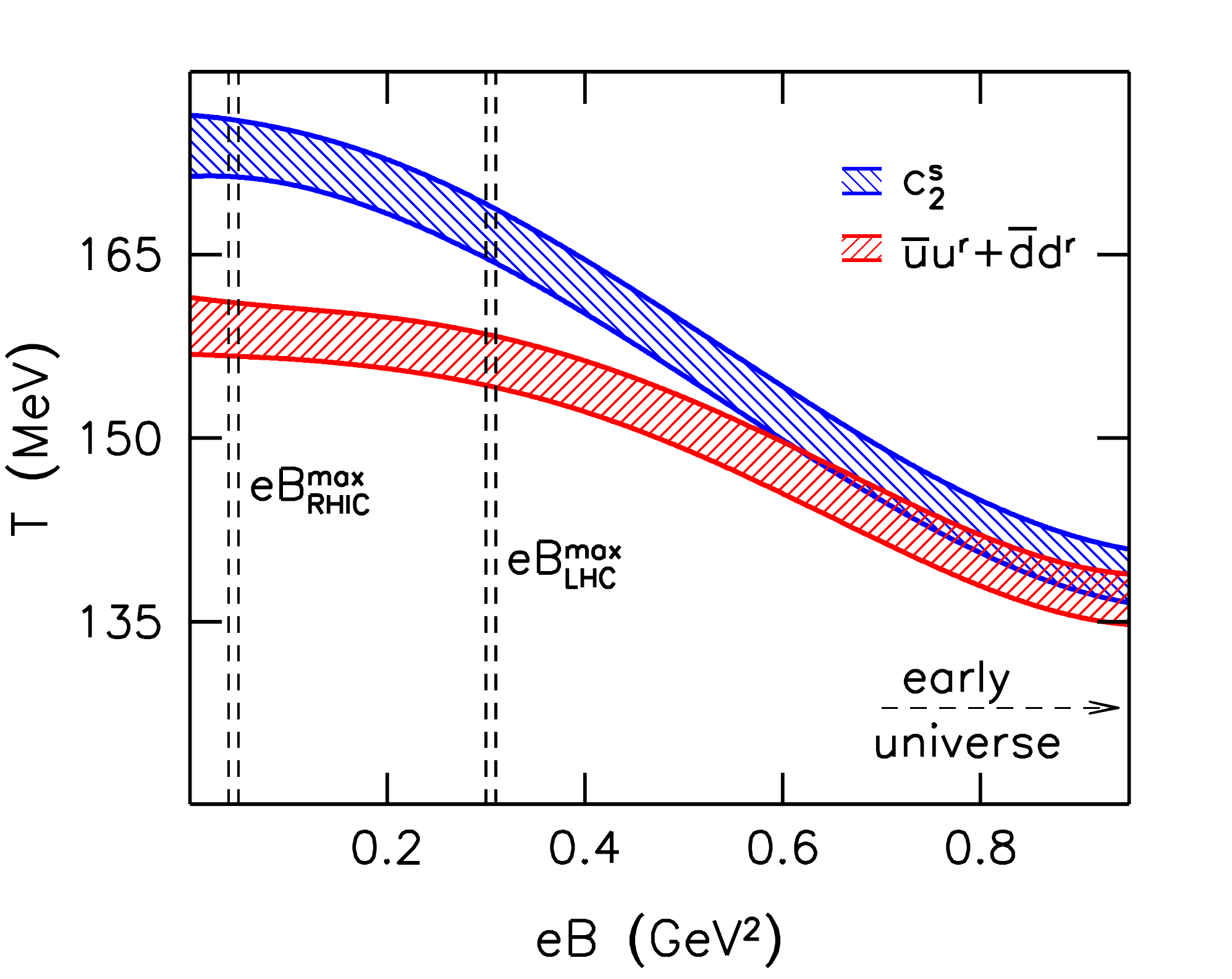}
\hspace{0.05\textwidth}
\includegraphics[height=0.2\textheight]{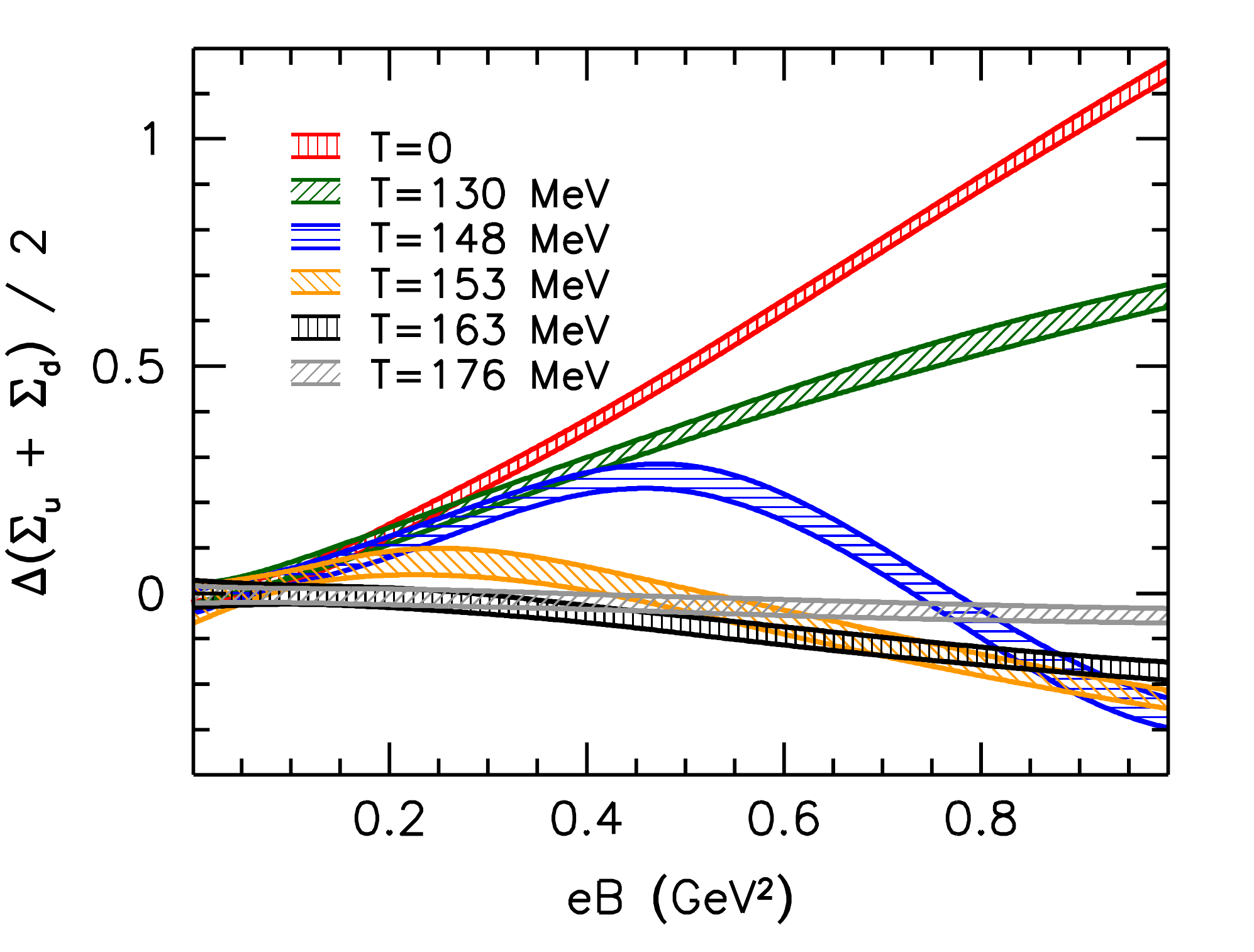}
\vspace{-5mm}
\end{center}
\caption{
\textit{Left:}
The inflection point of the chiral condensate and of the strange susceptibility.
The cross-over temperature decreases with a magnetic field, independently of
its definition \cite{Bali:2011qj}.
\textit{Right:}
The behaviour of the chiral condensate as a function of the magnetic field at
various temperatures \cite{Bali:2012zg}. The magnetic catalysis at low $T$
turns into a non-monotonic behaviour in transition region.
}
\label{fig:finB}
\end{figure}

We finally discuss the effect of a magnetic field. The strong magnetic fields
in non-central heavy ion collisions \cite{Skokov:2009qp} motivated the study
of the QCD transition in the presence of an electromagnetic background. 
A $B$ field is easily implemented on the lattice
\cite{Cea:2007yv,Ilgenfritz:2012fw}.  Based on models (e.g. PNJL
\cite{Gatto:2010pt}) and exploratory lattice simulations \cite{Cea:2007yv} the
expectation was to see an increase in $T_c$ with growing magnetic field.  As a
joint effort of the Budapest, Regensburg and Wuppertal lattice groups $T_c$ was
calculated with physical quark masses in the continuum limit.  $T_c$ was shown
to decrease with $B$ (see Fig.~\ref{fig:finB}/left), while the
transition was still a crossover \cite{Bali:2011qj}. To show the role played
by the quark mass, an auxiliary data set was
taken with higher hadron masses. Then the trend was reversed, and the
earlier (apparently contradicting) lattice result \cite{Cea:2007yv} was reproduced. To explain
the behaviour seen in the PNJL model the same group calculated the chiral
condensate at $B>0$ and $T>0$, and highlighted the differences between
lattice data and the model's assumptions \cite{Bali:2012zg} (see
Fig.~\ref{fig:finB}/right).

\section{Fluctuations of conserved charges}

Correlations and fluctuations of conserved charges have been proposed long ago
to signal the transition \cite{Jeon:2000wg,Asakawa:2000wh}.  At high
temperatures fluctuations are expected to approach the ideal gas limit. On the
other hand, in the low-temperature phase they are expected to be small since
quarks are confined and the only states with non-zero quark number have large
masses. Agreement with the Hadron Resonance Gas (HRG) model predictions is
expected in this phase.

In the past year both the Wuppertal-Budapest \cite{Borsanyi:2011sw} and HotQCD
\cite{Bazavov:2012jq} collaborations published results for the second order
fluctuations, i.e. the second derivatives of the free energy with respect to
chemical potentials of various conserved charges, like baryon number, electric
charge or strangeness.

The lattice results are always normalized to volume and the respective
power of temperature. E.g. the diagonal charge and the off-diagonal charge-baryon correlators are defined by
\begin{equation}
\chi_{2}^Q=\frac{1}{VT^3}\frac{\partial^2\log Z}{(\partial\mu_Q/T)^2}\,,\qquad
\chi_{11}^{BQ}=\frac{1}{VT^3}\frac{\partial^2\log Z}{(\partial\mu_Q/T)(\partial\mu_B/T)}\,.
\end{equation}

\begin{figure}[htbp]
\begin{center}
\includegraphics[width=0.45\textwidth]{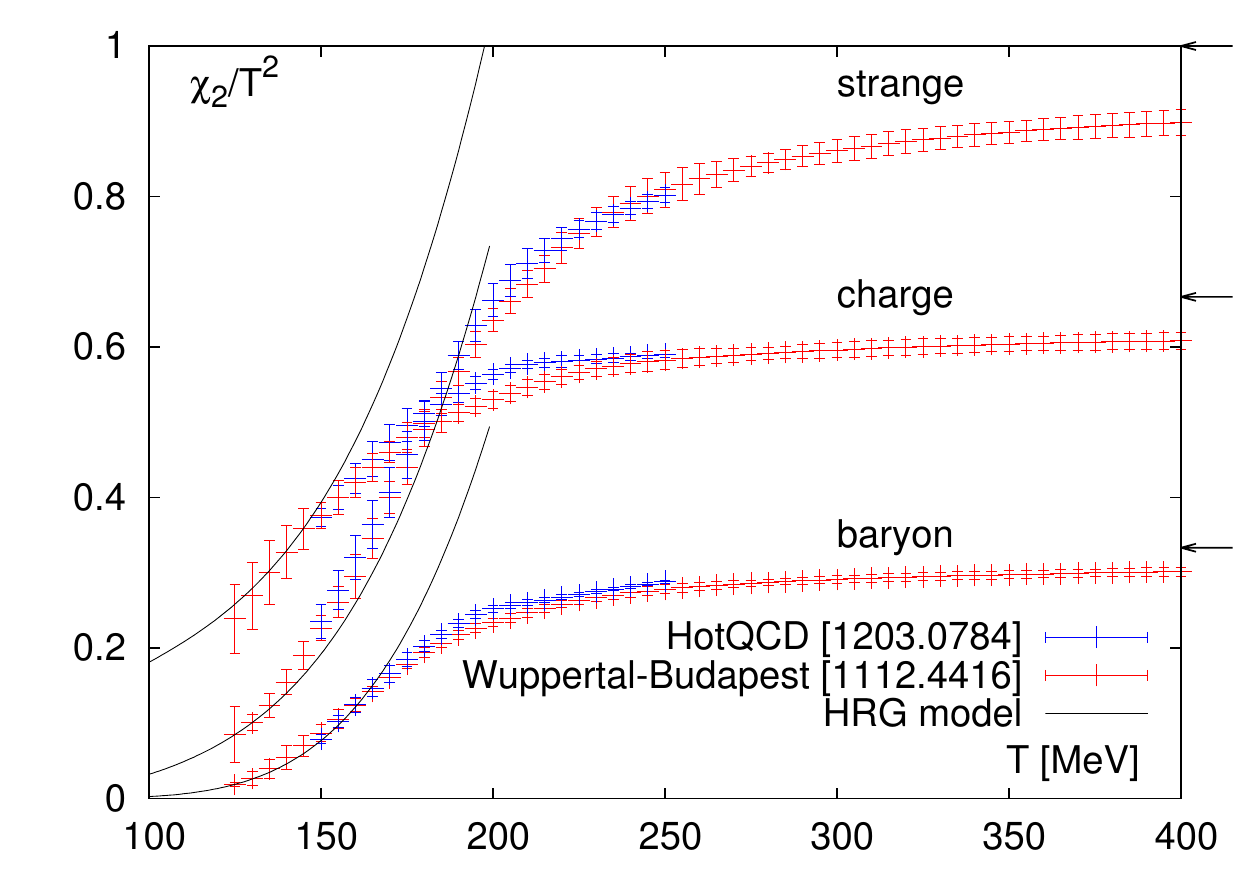}
\hspace{0.05\textwidth}
\includegraphics[width=0.45\textwidth]{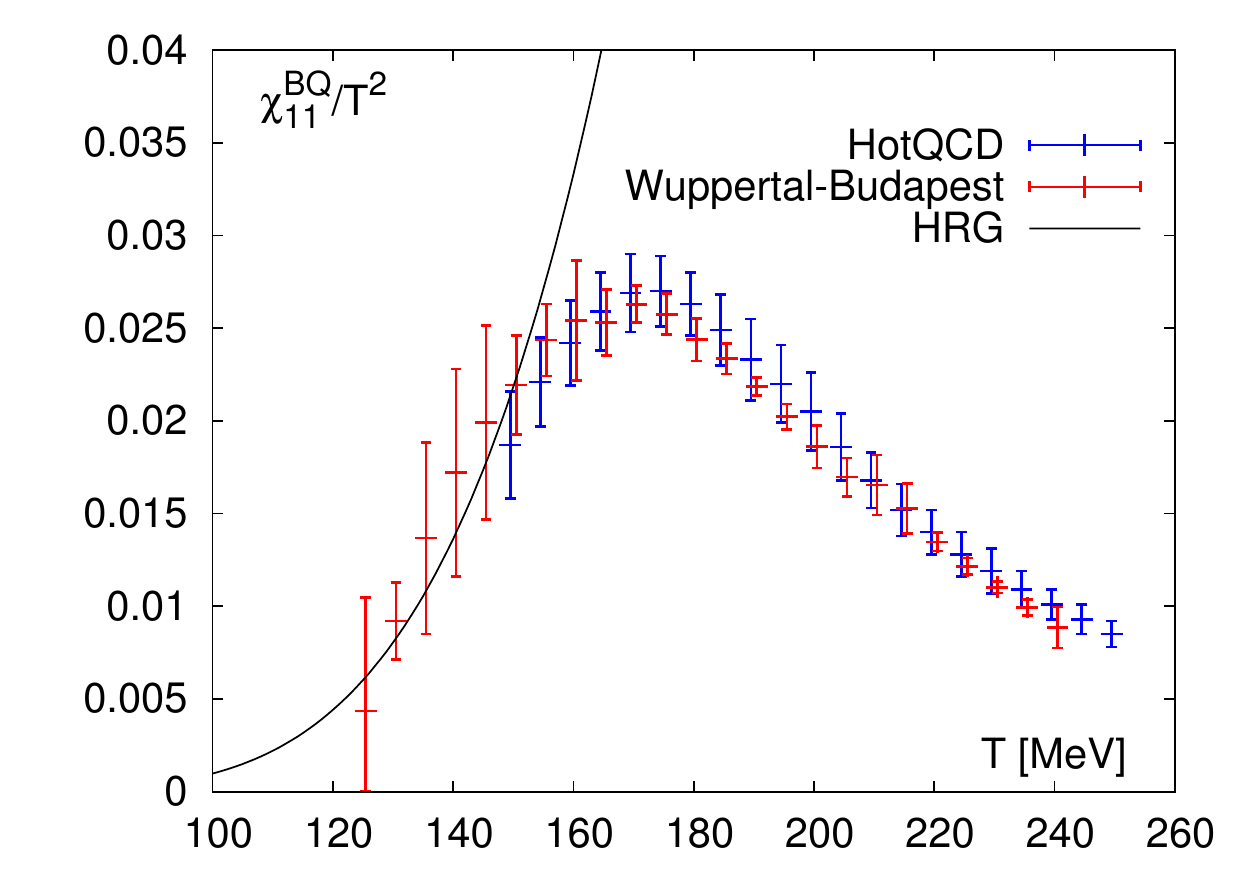}
\end{center}
\caption{
Diagonal and off-diagonal correlators of conserved charges as a function of
temperature. Only continuum extrapolated lattice data are shown.
There is a good agreement between HotQCD \cite{Bazavov:2012jq} and 
Wuppertal-Budapest \cite{Borsanyi:2011sw} data,
obtained using the \texttt{HISQ} and \texttt{2stout} actions, respectively. The broader
temperature region by the Wuppertal-Budapest collaboration allows a comparison
with the HRG model's prediciton, and was used to benchmark resummation
schemes at high temperature \cite{Andersen:2012wr}.
}
\label{fig:c2}
\end{figure}

The experimentally more interesting observables are ratios of these
derivatives, where the unknown volume factor cancels. To describe
the freeze-out parameters ($T$ and $\mu_B$) specific ratios have been
introduced and calculated (e.g.  skewness / mean of the net charge yield)
using the coarser data sets of the \texttt{HISQ} action ($N_t=6,8$) in
Ref.~\cite{Bazavov:2012jq}. 

\begin{figure}[tbp]
\begin{center}
\includegraphics[width=0.45\textwidth]{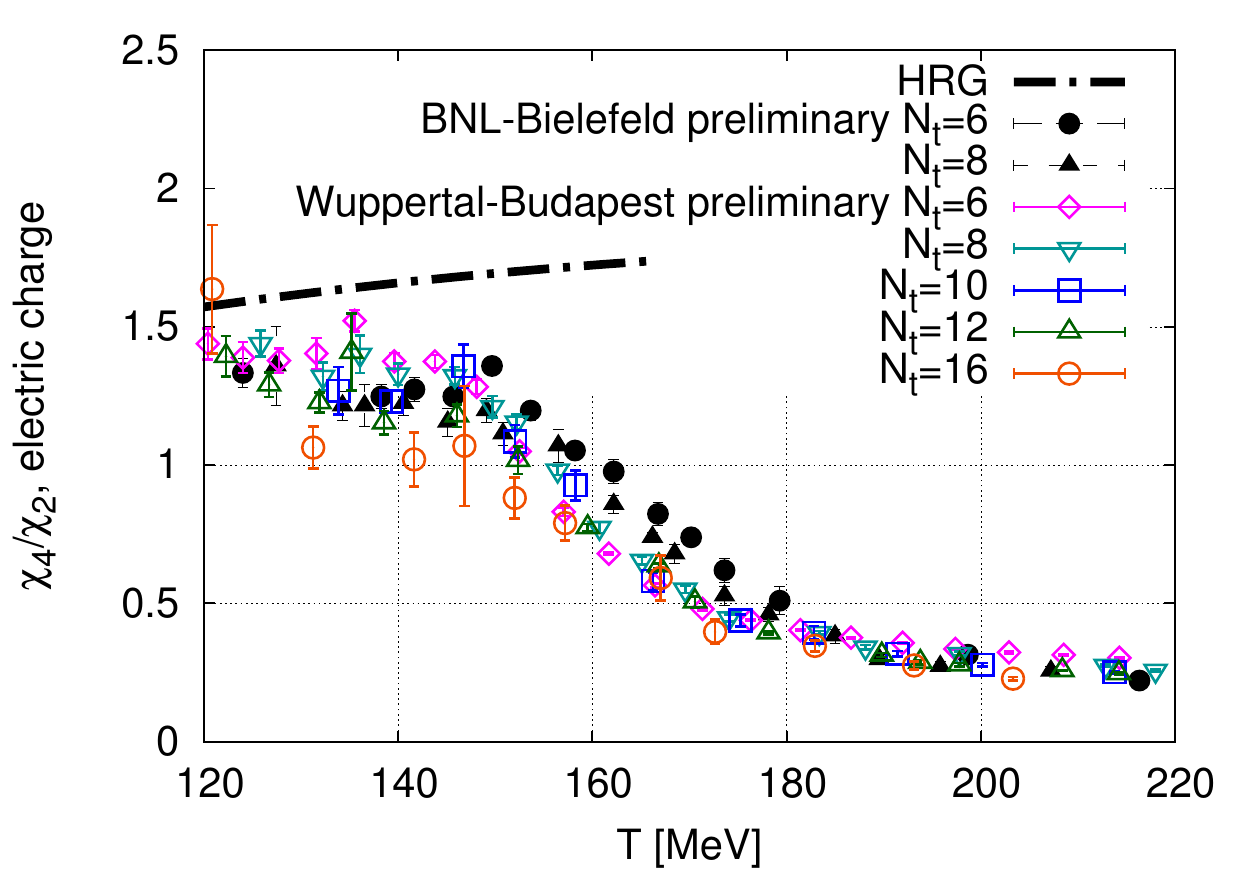}
\hspace{0.05\textwidth}
\includegraphics[width=0.45\textwidth]{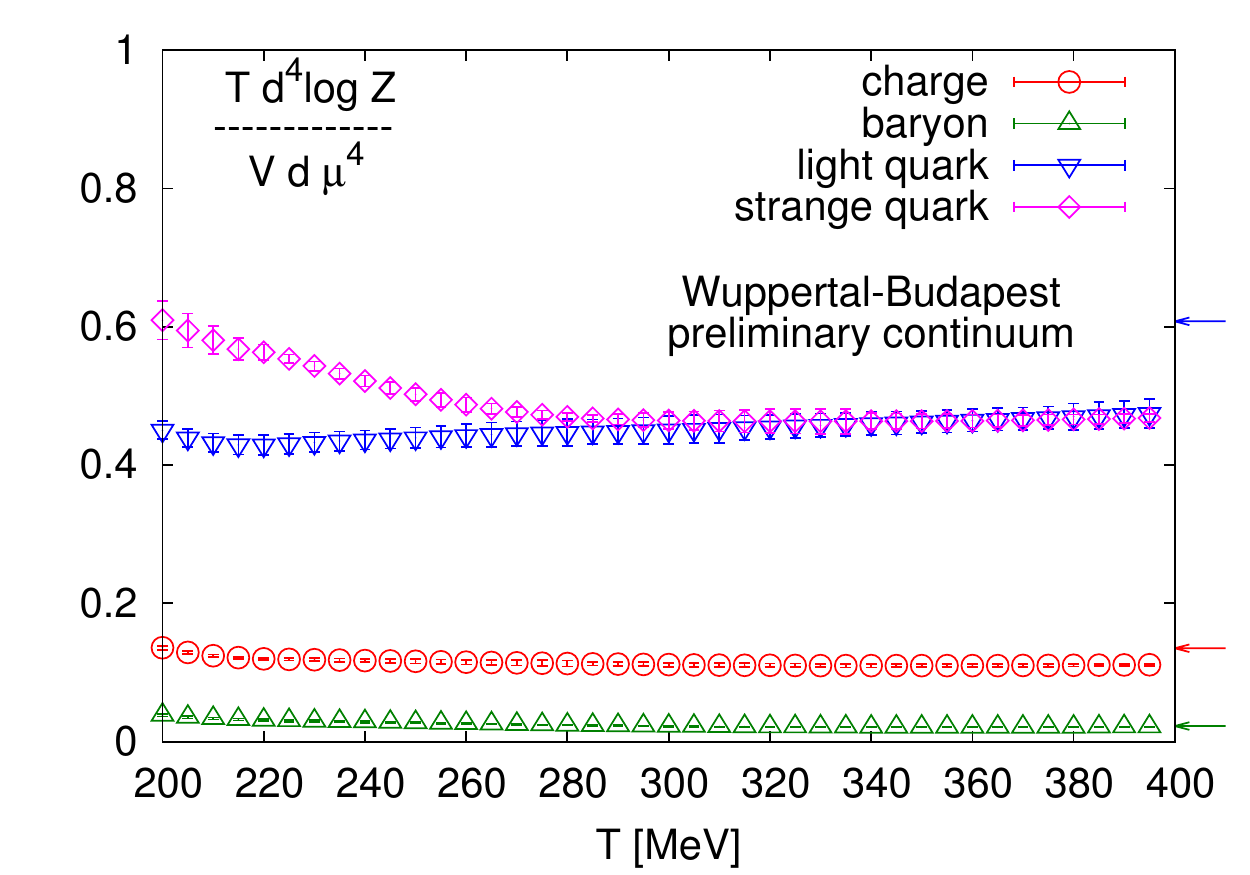}
\end{center}
\caption{
\textit{Left:} The normalized electric charge kurtosis from lattice QCD: the BNL-Bielefeld collaboration has used the \texttt{HISQ} action on $N_t=6$ and $8$ lattices \cite{ChristianQM12},
the Wuppertal-Budapest collaboration shows results at $N_t=6,8,10,12$ and $16$.
From the plot we see that the use of fine lattice is essential for a controlled
continuum limit. At present data indicates deviations from the Hadron Resonance
Gas model's prediction even below $T_c$.
\textit{Right:} At high temperatures the Wuppertal-Budapest collaboration
has attempted a preliminary continuum extrapolation for $\chi_4$ for electric
charge, strangeness and baryon number. The difference we see between light and
strange quark's behaviour indicates that the mass of the strange quark
is still noticeable up to about 260 MeV temperature.
}
\label{fig:kurt}
\end{figure}

The charge kurtosis $\times$ variance could also be used as a thermometer, and
was in the focus of C.~Schmidt's talk \cite{ChristianQM12}.  In
Fig.~\ref{fig:kurt} we plot the temperature dependence of $\chi^Q_4/\chi^Q_2$
at zero chemical potential from both the BNL-Bielefeld group and the
Wuppertal-Budapest collaboration. While we witness steady progress in the
analysis of experimental data towards the precise measurement of second and
higher order fluctuations \cite{MitchellQM12,McDonaldQM12,LuoQM12}, lattice
methods are increasingly successful in higher derivatives, which will finally
enable the extrapolation of the kurtosis data towards finite chemical
potential.

\section{Summary}

We discussed a selection of the recent developments in bulk thermodynamics,
focusing on those results with physical quark masses close to or in the
continuum limit. For technical reasons, all these results were obtained using the
staggered formalism. The independence of the results with respect to
the applied discretization scheme is an important consistency requirement of
lattice QCD, and to this end there have been recent checks with overlap
\cite{Borsanyi:2012xf}, domain-wall \cite{Bazavov:2012ja} and Wilson
\cite{Borsanyi:2012uq} fermions, with the compromise of using
heavy pions in these comparisons.

By today we know several aspects of the QCD transition through Euclidean-time
observables at zero and small chemical potentials. Now the challenge for lattice
QCD is to develop new techniques for extracting real-time physics (e.g.
transport coefficients) and to increase the accessible range in $\mu_B$,
and eventually, to map the phase diagram.

\end{document}